\documentclass[useAMS,usenatbib]{mn2e}

\def\msun{\hbox{M$_\odot$}}

\def\t4{\hbox{t$_{\rm 4}$}}

\newcommand{\au}{\rm AU}
\newcommand{\mmean}{\bar{m}}
\newcommand{\myr}{{\rm Myr}}
\newcommand{\pc}{\rm pc}
\newcommand{\mc}{M_{\rm c}}
\newcommand{\rdisc}{r_{\rm disc}}
\newcommand{\vstar}{v_{*}}
\newcommand{\rcore}{r_{\rm core}}
\newcommand{\kms}{{\rm km}\,{\rm s}^{-1}}
\newcommand{\fcross}{f_{\rm cross}}
\newcommand{\fincore}{f_{\rm in}}
\newcommand{\flm}{f_{\rm lm}}
\newcommand{\freq}{\Lambda_{\rm clear}}

\usepackage{graphicx}
\usepackage{natbib}
\usepackage{amssymb}
\usepackage{lscape}
\usepackage{longtable}
\usepackage{amsmath}
\title[Early Disc Accretion in Globular Clusters]{Early Disc Accretion as the Origin of Abundance Anomalies in Globular Clusters}
\author[N. Bastian et al.]{ N. Bastian$^{1}$\thanks{NB: N.J.Bastian@ljmu.ac.uk}, H.J.G.L.M. Lamers$^{2}$, S.E. de Mink$^{3,4}$\thanks{Einstein Fellow}, S.N. Longmore$^{1}$, \newauthor S.P. Goodwin$^{5}$, M. Gieles$^{6}$\\
$^{1}$ Astrophysics Research Institute, Liverpool John Moores University, 146 Brownlow Hill, Liverpool L3 5RF, UK\\
$^{2}$ Astronomical Institute "Anton Pannekoek", University of Amsterdam, Science Park 904, 1098 XH, Amsterdam, The Netherlands\\
$^{3}$ Observatories of the Carnegie Institution for Science, 813 Santa Barbara St, Pasadena, CA 91101, USA \\
$^{4}$ Cahill Center for Astrophysics, California Institute of Technology, Pasadena, CA 91125, USA\\
$^{5}$ Department of Physics and Astronomy, University of Sheffield, Sheffield S3 7RH, UK\\
$^{6}$ Department of Physics, University of Surrey, Guildford GU2 7XH, UK\\
}
\begin{document}

\date{Accepted XXX. Received XXX; in original form XXX}

\pagerange{\pageref{firstpage}--\pageref{lastpage}} \pubyear{2013}

\maketitle

\label{firstpage}

\begin{abstract}
Globular clusters (GCs), once thought to be well approximated as simple stellar populations (i.e. all stars having the same age and chemical abundance), are now known to host a variety of anomalies, such as multiple discrete (or spreads in) populations in colour-magnitude diagrams and abundance variations in light elements (e.g., Na, O, Al).  Multiple models have been put forward to explain the observed anomalies, although all have serious shortcomings (e.g., requiring a non-standard initial mass function of stars and GCs to have been  initially 10-100 times more massive than observed today).  These models also do not agree with observations of massive stellar clusters forming today, which do not display significant age spreads nor have gas/dust within the cluster. Here we present a model for the formation of GCs, where low mass pre-main sequence (PMS) stars accrete enriched material released from interacting massive binary and rapidly rotating stars onto their circumstellar discs, and ultimately onto the young stars. As was shown in previous studies, the accreted material matches the unusual abundances and patterns observed in GCs.  The proposed model does not require multiple generations of star-formation, conforms to known properties of massive clusters forming today, and solves the ``mass budget problem" without requiring GCs to have been significantly more massive at birth.  Potential caveats to the model as well as model predictions are discussed.

\end{abstract}


\section{Introduction}
\label{sec:intro}

Our understanding of globular clusters (GCs) and their formation has undergone a radical change in the past two decades.  First, it is now clear that while traditionally thought of as the quintessential simple stellar populations (i.e., all stars within a cluster have the same chemical abundances and age within some small tolerance), globular clusters host multiple stellar populations with spreads in He, many light elements (e.g., Na, O, Al) and even Fe in some cases (e.g., Gratton, Carretta, \& Bragaglia~2012).  Secondly, globular clusters, once thought to only be able to form in the special conditions present in the early Universe, are still forming today (e.g., Holtzman et al.~1992; Schweizer \& Seitzer~1998).  These clusters (knows as Young Massive Clusters - YMCs), with ages of  $<1$~Gyr, have masses and densities similar to, or significantly exceeding those of, GCs.

The origin of the multiple populations within clusters is still under debate and multiple models have been put forward, which are outlined below.  These models make specific predictions that can be tested against the wealth of information now available for GCs as well as for young massive clusters.  

\subsection{Observational Constraints}
\label{sec:constraints}

Here, following on Renzini~(2008), we summarise the main observations that need to be reproduced by any model attempting to explain the chemical anomalies within GCs. In particular, any model must:

\begin{enumerate}

\item Have a more centrally condensed enriched population compared to the unenriched population (e.g., Lardo et al.~2011).

\item Reproduce the observed Na-O anti-correlation as well as the range in values (e.g., Carretta et al.~2009a).

\item Have large spreads in Al, and relatively small spreads in Mg. While a few clusters do show a pronounced Al-Mg anti-correlation at high Al values (large Mg spread towards low Mg values), the majority do not (see \S~\ref{sec:mgal}).

\item Create main-sequences that show spreads or multiple quantised sequences, depending on the cluster and the filter range used.  If spreads are seen in the optical, they are likely to be caused by differing He abundances (up to $\delta Y=0.13$ - Piotto et al.~2007, although often significantly smaller $\delta Y \le 0.03$).  He spreads may also affect the morphology of the Horizontal Branch.  If main-sequence spreads are observed in ultraviolet filters, this is likely due to the anti-correlation of C and N (Sbordone et al.~2011).

\item Account for the observation that stars which appear to be He-enriched, also show variations in Na, O, Al, and sometimes Mg (i.e. the enrichment of various elements are correlated).

\item Show relatively small spreads in Li ($<0.3$~dex - e.g., Monaco et al.~2012), and if spreads are seen then Li should be anti-correlated with Na and correlated with O (Pasquini et al.~2005).

\item Explain that the sum of C+N+O appears to be constant amongst cluster stars within a factor of 2-3 (c.f. Decressin et al.~2009).

\item Display spreads in light elements (Al, Na, O), hence the gas needed for the enrichment cannot be fully mixed.

\item Produce relatively small (or no) Fe spreads within the cluster (e.g., Carretta et al.~2009b)\footnote{The origin of GCs with significant Fe spreads, e.g., $\omega$-Cen, is currently under debate.  They may share the same formation history as other GCs, or they may be the nuclei of dwarf galaxies that have been accreted onto the Galaxy.  In the later case, they would likely have a significantly different formation chanel.}.

\item Satisfy the mass budget constraints within the clusters, that the enriched population make up $30-70$\% of the current stars within the clusters (e.g., Carretta et al.~2009a).  We will refer to this as the ``internal mass budget".

\item Satisfy the mass budget constraints imposed by observations of the Fornax Dwarf galaxy (Larsen et al.~2012), where the current GCs cannot have been more than five times more massive at birth than they are now.  We will refer to this as the ``external mass budget".

\item Explain that while there are common trends observed in GCs (i.e. the Na-O anti-correlation and implied He-spreads) the specifics of each cluster are often unique (e.g., Renzini 2008).  Hence, the model must explain the commonality as well as allow for differences observed between clusters.

\item Explain why enriched stars make up only $\sim3$\% of stars in the halo of the Galaxy (e.g., Martell et al.~2011), although this observation is often linked with item (i) (e.g., Vesperini et al.~2010).

\end{enumerate}

If the mechanism that created the multiple populations in GCs is operating in young massive clusters observed today, then the following additional constraints must be satisfied (otherwise the model must explain why it is not observed today):

\begin{enumerate}
\setcounter{enumi}{13}

\item Clusters with current masses up to $10^5$\msun\ and ages of $<300$~Myr cannot display age spreads, or distinct bursts with separations/durations $>30$~Myr (Bastian \& Silva-Villa~2013).

\item For clusters with masses up to $10^7$\msun, they cannot have continuous star-formation within them for more than $\sim20$~Myr (Bastian et al.~2013), unless the stellar IMF is truncated above $15$\msun.  Discrete bursts of star formation are not favoured by current observations, but are not formally ruled out due to the finite sampling of ages and masses of the clusters studied to date.

\item Clusters with masses up to $10^5$\msun\ cannot have significant amounts of dense gas left over from the star-formation process for more than $3$~Myr (see above references).

\item To date, no study has found significant amounts (i.e. $\sim10$\% of the cluster stellar mass) of gas within clusters in ionised, atomic (warm), or molecular (cold) states for clusters with ages in excess of $3-5$~Myr.  Further observations are required to quantify the precise limits, but for ionised gas (10,000~K), current limits are $<100$~\msun\ even within the most massive clusters (Bastian et al.~2013).


\end{enumerate}

Ongoing spectroscopic work (Cabrera-Ziri et al. in prep) will place tight constraints on the allowed instantaneous ``secondary burst" within clusters, i.e. is a second burst seen in any YMCs with ages between $10$ and $\sim500$~Myr.  This follows on the above constraints, but does not require that we catch the cluster undergoing the burst (i.e. we can infer if a secondary burst happened $10-300$~Myr previously).  We also note that the young ($400-500$~Myr) massive clusters, W3 and W30 in the merger remnant NGC~7252, ($8 \times 10^7$\msun\ - Maraston et al. 2004 and $1.7\times10^7$\msun\ - Bastian et al.~2006) are both consistent with a single burst of star formation (Schweizer \& Seitzer~1998).  Of all young massive clusters observed to date, only nuclear clusters (i.e., those in the centres of galaxies) show evidence for multiple episodes of star formation (e.g., Rossa et al.~2006; Seth et al.~2008), although due to the depths of their potential wells and gas feeding timescales, they presumably also have significant [Fe/H] spreads.

\subsection{Previous Models and Potential Drawbacks}
\label{sec:previous_models}

The observed anomalies in GCs are not seen in young metal-rich open clusters or in the majority of field stars in the halo of the Galaxy (e.g., Martell et al. 2011; Gratton et al.~2012).  This has lead to the suggestion that the abundance trends are due to the unique formation environment of GCs in the early Universe and that GCs, potentially due to their large masses, have undergone multiple star forming events (e.g., D'Ercole et al.~2008; Conroy~2012).   It has also been suggested that stars are able to accrete material from the ejecta of higher mass main-sequence stars (D'Antona, Gratton \& Chieffi 1983), hence the observed abundance patterns would simply reflect surface contamination.  However, this model is not feasible as evolved red giant branch stars (which have mixed their surface and lower layers through convection) show the same Na-O anti-correlations as main-sequence stars (Cohen, Briley \& Stetson~2002), indicating that the abundance anomalies are present throughout the star (or at least as deep as the giant branch dried-up events).

More recent models invoke the formation of a second population of stars from the processed material of specific stars from the first population.  Only certain stars are able to produce He enriched material that also displays the observed chemical abundance (anti-)correlations.  The stellar sources that have been proposed are: Asymptotic Giant Branch (AGB) stars (e.g., D'Ercole et al.~2008), fast rotating massive stars (also known as spin-stars - e.g., Decressin et al.~2007), and massive stars in interacting binary systems (de Mink et al.~2009 - hereafter dM09).  

The models that invoke multiple star-formation events within a single cluster have achieved several notable successes, explaining many of the observed chemical properties and naturally predicting that the enriched population should be more centrally concentrated within GCs. However, such models all suffer from several drawbacks. Firstly, if one assumes that GCs formed in a similar way to young massive clusters, and if multiple episodes of star formation are required to explain the abundance patterns, then a natural consequence is that we should see young clusters with ongoing star formation or extended star formation histories.  However, all young massive clusters ($10^4 - 10^8$\msun) observed to date appear to be well represented by a single burst of star formation with little or no age spread (e.g., Schweizer \& Seitzer~1998; Kudryavtseva et al. 2012; Bastian \& Silva-Villa 2013; Bastian et al.~2013; Cabrera-Ziri et al. in prep.; see \S~\ref{sec:intermediate} for a discussion of the $1-2$~Gyr old massive clusters in the LMC/SMC).

Secondly, massive clusters appear to be free of cold/dense gas/dust from young ($<3$~Myr) ages (e.g., Muno et al.~2006; Sana et al.~2010; Campbell et al.~2010; Seale et al.~2012), suggesting that they cannot retain the low velocity outflows required to form the second generation.  It has been suggested by Conroy \& Spergel~(2011), that the enriched gas within the clusters is kept warm by Lyman continuum photons from B and A-stars, which keeps it from collapsing to form stars.  However, in a survey of 129 YMCs ($10^4 - 10^8$\msun, $10-1000$~Myr) no clusters were found with ionised gas associated with them (with the exception of three clusters hosting PNe - Bastian et al.~2013), down to limits of $<100$\msun\ of ionised gas.  Future studies should quantify the amount of gas in warm (atomic) and cold (molecular) states, if any, within massive young clusters. 

Thirdly, the models require that the first generation of stars within the cluster was significantly more massive than observed today (by factors of 10 to 100) in order to create enough material from the binaries, spin-stars, or AGB stars to form the number of second generation stars that are observed (e.g., Conroy~2012).  This extra-mass must be retained by the cluster for a few Myr to tens/hundreds of Myrs, in order to be used to form the second generation, then lost rapidly (e.g., Conroy~2012).  However, a comparison between the number of metal-poor stars in the field and in GCs in the Fornax Dwarf galaxy, places a strict upper limit of a factor of 5 in mass to the difference between the initial cluster mass and that observed (Larsen, Strader, Brodie 2012), in conflict with model predictions.  One potential solution to this problem is if the second generation had a radically different stellar initial mass function, in particular truncated at $\sim0.8$\msun, i.e. all stars formed in the second generation are still alive today (e.g., D'Antona et al.~2013)\footnote{D'Antona et al.~(2013) have discussed a possible mis-match between the metallicity scale of the GCs and field stars in Fornax.  However, given that Larsen et al.~(2012) group all low metallicity $[Fe/H] < -2$, together, any systematic offset is unlikely to significantly affect their results.}.  The Larsen et al. constraint is an upper limit to the amount of mass that could have been lost from GCs, because the study implicitly assumes that no GCs have been completely destroyed.  If one adopts the standard paradigm of the evolution of GC systems, that for each surviving GC there are $\sim10$ lower mass clusters that have been completely destroyed (e.g., Fall \& Zhang~2011; Lamers et al.~2010), the present day GCs could not have been any more massive than they are now.

Fourthly, observations show that the chemical anomalies within clusters show spreads, i.e. are not discrete (e.g., Carretta et al.~2009a).  While some of the spread can be due to measurement errors and some bi-modality may exist (e.g., Villanova \& Geisler~2011), recent measurements show that the spread is significantly larger than the estimated errors (e.g., Carretta et al.~2013).  For models that invoke the formation of a second generation of stars from a mixture of pristine and chemically enriched gas, it is difficult to not have the gas well mixed.  For example, if the enriched and pristine gas is kept from forming stars by being heated (either warm or ionised) in order to collect enough pristine and enriched material to form the second generation (at T=$100 - 1,000$~K - Conroy \& Spergel~2011), the mixing time, $t_{\rm mix} = R/v$ is $0.3-1.5$~Myr (assuming R=1pc and $v=0.7 - 3$~km\,s$^{-1}$).  Hence, any gas present in the cluster would quickly mix leading to a chemically homogeneous gas cloud. Unless extended star-formation histories are invoked (where stars could form out of gas with different chemical abundances), these models would predict only two populations, either 'pristine' or 'enriched' stars.

Finally, the models proposed so far that invoke the formation of a second population all have difficulties reproducing the full range of abundance trends observed in GC.  For example, the AGB model does not predict the Na-O anti-correlation (the ejecta from AGB stars are expected to show a Na-O correlation), so pristine material must be brought in just as the second generation is forming.  Additionally, if high-mass stars are the origin of the enriched material, the second generation is expected to be depleted in Li.  Hence, in this model some pristine material is also required (Decressin et al.~2007; dM09).  

Due to these shortcomings, alternative models for the chemical anomalies in GCs should be explored.  In particular, models that are motivated by observations of young globular clusters (e.g., young massive clusters with masses and densities equal to, or exceeding that of, GCs) may provide a new and unique perspective into this perplexing problem.  It is with this in mind that we suggest a new model for the origin of chemical anomalies in GCs.

We note that the different populations within GCs are often refereed to as ``1st and 2nd generations", with the ``2nd generation" being the stars that have been enriched.  This implies that distinct star forming events have taken place, whereas that may not be the case.  Instead, we will refer to the multiple populations as unenriched and enriched.

This paper is organised as follows.  In \S~\ref{sec:model} we introduce the ``early disc accretion" model for the observed chemical anomalies observed in GCs.  In \S~\ref{sec:assumptions} we discuss the assumptions inherent in the model and potential caveats that require further investigation, while in \S~\ref{sec:predictions} we outline specific predictions of the model.  Our conclusions are given in \S~\ref{sec:conclusions}.


\section{The Early Disc Accretion Model}
\label{sec:model}

\subsection{The Basics of the Model}
\label{sec:basics}

Our model envisions a gas-free cluster (i.e. all primordial material not used in star-formation has been removed - see \S~\ref{sec:initial}) with a standard stellar population  and a Kroupa~(2001) initial mass function with the high-mass stars concentrated towards the cluster core (i.e. mass segregated\footnote{This mass segregation need not be primordial in nature, as it could happen dynamically on a timescale of less than a few Myr, if the cluster formed from the merging of sub-clumps (McMillan et al.~2007; Allison et al.~2009).} - see Portegies Zwart, McMillan, \& Gieles~2010).  Chemically enriched material, ejected from high mass interacting binaries or spin-stars (see also \S~\ref{sec:nature} for alternative enrichment sources) is then accreted onto the circumstellar discs and ultimately onto low mass PMS stars that were formed in the {\em same generation}.  Higher mass main sequence stars stop accreting earlier, allowing only the low mass stars to accrete the enriched material.   The process would only strongly affect the stars that pass through the cluster core ($<50$\% for an isotropic velocity distribution, see \S~\ref{sec:distribution} for how we define the core, although we note that the exact definition does not affect our results), such that about half of the low mass stars have normal abundances, because they did not cross the core.   The proposed model is expected to apply to the same types of GCs as the previously proposed models, i.e., those clusters with low (or no) Fe spreads within their stellar populations (see \S~\ref{sec:fe_spreads}).

\subsection{``Tail-End" Accretion}
\label{sec:tail_end}

The PMS duration is related to the stellar mass, with lower mass stars remaining in the PMS phase for a longer time.  For example, stars of 2, 1 and 0.5 \msun\ have PMS lifetimes of approximately 6, 25, and 90 Myr, respectively (Siess, Dufour, \& Forestini~2000; Baraffe et al~2002). It is likely that once stars reach the main sequence they cannot accrete a significant amount of new material due to radiative feedback and stellar winds, allowing lower mass stars to accrete for longer periods.  The accretion of enriched material from the ejecta of high-mass stars has previously been suggested in young star forming regions, such as Orion, to explain abundance anomalies there, and is known as ``tail-end" accretion (Throop \& Bally~2008).  ``Tail end" accretion refers to accretion during the later stages of a disc's lifetime, after the main phase of star formation is over, but before the disc is dissipated away. Normally, the abundance of a non-evolved star is thought to be set during its formation, however ``tail end" accretion offers a mechanism to alter the star's final abundance, and to some degree, its final mass.

While the concept of "tail-end" accretion was developed with supernovae pollution in mind, the low velocity ejected material of massive binaries (or spin stars) appears more plausible (e.g., Chevalier~2000; see also \S~\ref{sec:fe}).

Due to the high velocity dispersion of massive clusters ($15-45$~km\,s$^{-1}$; e.g., Bastian et al.~2006), direct accretion onto the surfaces of the young stars through gravitational focussing is unlikely (Bondi \& Hoyle~1944).  However, if a star has a circumstellar disc of material, the disc can entrain material as it moves through the interstellar medium (ISM), and this material eventually accretes onto the star (Chevalier~2000; Moeckel \& Throop~2009).  Unlike gravitational focussing which becomes less efficient at high relative velocities between the star/disc and the ambient gas, the efficiency of the gas capture by discs is proportional to the relative velocity.  We will refer to this as the "disc sweep-up model".

Since the PMS phase is limited to the initial part of a starÕs lifetime, we focus on potential polluters that can operate in the first $\sim5-10$ Myr of a clusterÕs existence.  Observations of nearby star-forming regions suggest that the majority of high-mass stars are in binary systems that will interact at some stage of their evolution (Sana et al.~2012), hence we focus on this type of polluter system.  Interacting massive binaries have been shown to slowly eject a large portion of the primary star's envelope that is enriched in He, and displays all/most of the known abundance (anti-)correlations (dM09 - see also \S~\ref{sec:abundances}). This material, which can make up to $\sim13$\% of the initial mass contained in stars of the cluster (dM09), can then be entrained by the circumstellar discs of low mass stars, enriching them and causing the star to grow in mass.  Note that higher mass stars, $>2$\msun, are already on the main sequence, where their radiation, stellar winds and lack of a circumstellar disc prevents them from accreting any new material.

The enriched material from interacting binaries is preferentially ejected in a disc geometry, with expansion speeds of $10-20$~km\,s$^{-1}$, and densities $>10^5$~cm$^{-3}$ (e.g., Smith, Bally, \& Walawender~2007).  These extreme densities are likely to limit the effects of photoionisation and SNe on the expanding discs (see also \S~\ref{sec:ram}).  Additionally, the dust temperatures in such ejecta are $85-190$~K (Smith et al.~2013 - consistent with the thin disc geometry), allowing the material to accrete onto the circumstellar discs of low mass stars.

The material that accretes onto the host pre-main sequence starÕs disc will have a lower local specific angular momentum than that of the main disc, which will drive a flow of material through the disc towards the central star (Lin \& Pringle~1990).  This is likely to have two effects.  Firstly, it may lead to a higher accretion rate onto the star.  Secondly, it may halt viscous expansion of the disc and add mass to the disc, the latter effect may lead to longer disc lifetimes.

\subsection{The Distribution of the Accreting Stars}
\label{sec:distribution}

We have assumed that the cluster is mass segregated, with the high mass stars concentrated towards the centre.  In this section we look at the spatial distribution of the accreting stars.  In particular, we look at the fraction of low-mass stars that enter the cluster core and the fraction of time that they spend there.

\begin{figure}
\includegraphics[width=8.5cm]{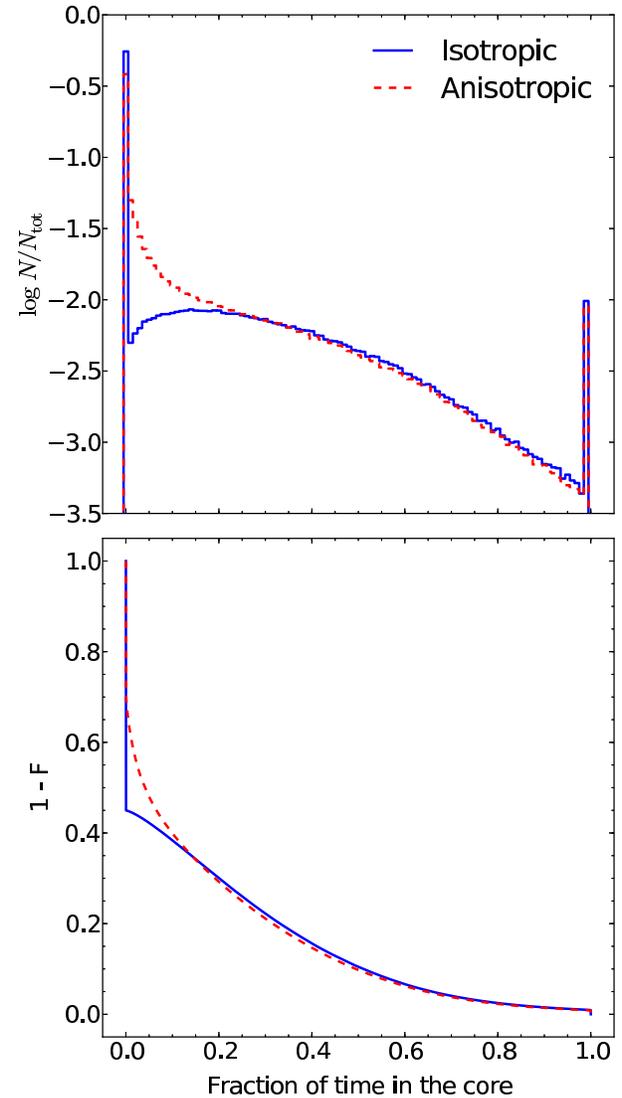}
\caption{The distribution of time spent by stars in the core of a star cluster. The lines show the fraction of stars that spend a given amount of time in the core ({\bf top:} histogram of the relative distribution; {\bf bottom}: the fraction of time spent in the core, or more). This is based on stellar orbits in the isochrone model of a globular cluster (Henon~1959), for different assumptions about the types of orbits present. }
\label{fig:distribution1}
\end{figure}

To compute how often stars cross the cluster core and with what velocity, we use the isochrone model (Henon~1959). The potential of this model is
\begin{equation}\Phi ( r )= -GM/(r_0 + (r_{0}^{2} + r^{2})^{0.5}). \end{equation}
where G is the gravitational constant, M is the cluster mass, r$_{0}$ is a scale radius (core radius) and r is the distance to the centre. The half-mass radius of this model is $r_h = 3.06 \times r_0$. The reason we choose this model is two-fold: orbits in this potential are analytic and the density profile resembles what is observed for young massive clusters in the LMC (Elson, Fall, \& Freeman~1987): a constant density core with a r$^{-4}$ fall off in three dimensions outside the core, corresponding to a r$^{-3}$ fall off the surface density profile.  At any given time, the mass fraction within the core is 12.1\%.  In Fig.~\ref{fig:distribution1} we show the fraction of time spent in the core for stars in this potential, for different assumptions on the types of orbits present.

We assume that the high mass stars are the source of the enriched material and that they are all contained within the core radius of the cluster.  Additionally, we adopt the disc Òsweep upÓ model, where the disc accretion is proportional to the stellar velocity and time spent in the core.  This follows from the orbits presented in Fig.~\ref{fig:distribution1}.  For an isotropic velocity distribution, the median time spent for the stars that enter the core, is ~1/3 of their time.  However, only 45\% of stars ever enter the core. These are the stars that accrete enriched material and may account for the quantised enrichment of He inferred from multiple discrete main sequences of stars observed in the many massive globular clusters. If clusters form through the merger of sub-clumps, an orbital distribution that favours radial orbits in the outer parts of the cluster is expected. This does not affect our conclusions, as the isotropic and radially anisotropic orbital distributions agree within ~10\% for stars that spend more than 10\% of their time in the core.

\begin{figure}
\includegraphics[width=8.5cm]{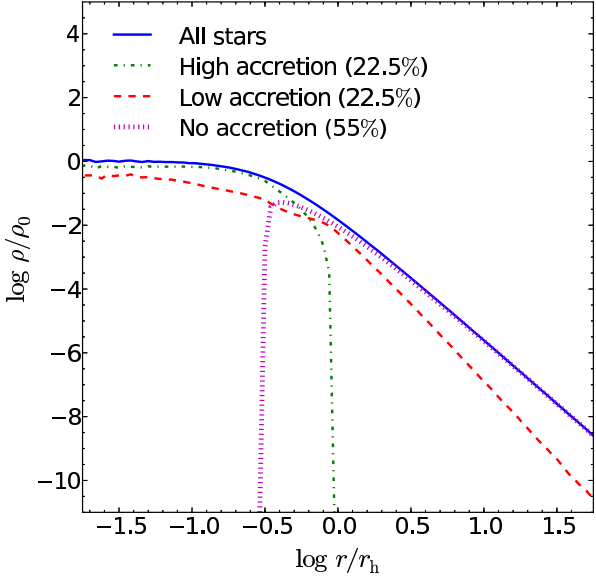}
\caption{The three dimensional density distribution of stars for a cluster with an isotropic velocity distribution.  The stars are separated into different classes based on their average accretion.  ``No accretion" stars are those that never enter the core, while low and high accretion stars are separated based on the median accretion rate for stars that enter the core.  Note that the high accretion rate stars are nearly exclusively in the core, whereas the non-accreting stars are never found in the core (by model construction).  The half mass radius is labelled r$_{\rm h}$ while the central density is $\rho_{0}$}
\label{fig:distribution2}
\end{figure}

We split the stars into three classes; stars that never enter the core, hence do not accrete, and for the stars that will accrete, we make an arbitrary split at the median accretion rate (i.e. high and low accretion stars).  The three dimensional density of these different classes is shown in Fig.~\ref{fig:distribution2}.  The high accretion stars are concentrated, nearly entirely, in the core of the cluster, whereas the low accreting stars follow (to first order) the underlying total distribution of the cluster.  Non-accreting stars, by construction, never enter the core.   

We also investigated cases where the stars in the cluster have a radially anisotropic velocity distribution.   For this case we used the Osipkov-Merritt description (Osipkov~1979; Merritt~1985). The orbits are isotropic in the inner parts and the anisotropy parameter increase gradually to fully radial orbits in the outer parts. We adopted an anisotropy radius equal to the half-mass radius. We note that even if the orbits of stars are strongly radially anisotropic, the conclusions remain largely unchanged, with a large fraction of stars spending little time in the core ($<10$\% of their orbits) and being within 10\% of the isotropic case for the rest of the stars.

Only the low-mass stars that spend a significant amount of time within the cluster core (where the high mass binaries reside) accrete enough of the processed material to show large He enhancements, potentially leading to ÔquantisedÕ abundances.  This issue is further addressed in \S~\ref{sec:nature}. 

\subsection{The Accretion Rate}
\label{sec:accretion_rate}

To estimate the frequency with which the core gets cleared by discs, $\freq$, we need to know: the fraction of the total number of stars that crosses the core ($\fcross$) and the fraction of time these stars spend in the core ($\fincore$). The distribution of the latter quantity is shown in Fig.~\ref{fig:distribution1}  for orbits in the isochrone model. This figure shows that both for the isotropic and the radially anisotropic velocity distribution, a fraction $\fcross\simeq 0.5$ of the stars cross the core during their orbit through the cluster (this is found from  the $y$ value at $x=0$ of the bottom panel of Fig.~\ref{fig:distribution1}). These stars spend on average a fraction $\fincore\simeq 0.33$ of their time in the core (this can not be read of from Fig.~\ref{fig:distribution1}, but is found by computing the mean of the individual values). Low-mass stars ($< 1\,\msun$) represent a fraction $\flm \simeq 0.9$ of the total number of a stars, which in turn can be found from the total mass of the cluster $\mc$ divided by the mean mass of low-mass stars ($\mmean\simeq0.4\,\msun$). The disc clearing frequency $\freq$ is proportional to the stellar velocities $\vstar$ and the area of the disc ($= 1/2 \pi\rdisc^2$ - where the 1/2 is a correction factor to account for the angle of the disc relative to the direction of travel) and inversely proportional to the volume of the core ($V = (4/3)\pi\rcore^3$).

For typical cluster values we then find that
\begin{multline}
\freq\simeq 0.75\, \myr^{-1}\frac{\fcross}{0.5}\frac{\fincore}{0.33}\frac{\flm}{0.9}\frac{\mc}{10^6\,\msun}\frac{0.4\,\msun}{\mmean} \\ \times \frac{\vstar}{20\,\kms}\left(\frac{\rdisc}{100\,\au}\right)^2\left(\frac{\rcore}{1\,\pc}\right)^{-3}.
\end{multline}
This frequency implies that the core gets cleared of gas once every $\sim1.3$\,\myr.



The above estimate assumes that the discs had a constant size, 100~AU, throughout the period of accretion.  This is unlikely to be the case, as discs around low-mass stars are seen to be significantly larger than this at young ages ($\sim1000$~AU - e.g., Lada et al.~2001), which may become truncated due to dynamical encounters and/or photoionisation.  Hence, the disc clearing is expected to be more rapid at earlier times than the estimate above, and decrease to the approximation above, over a timescale set by the disc truncation timescale.  This is further discussed in \S~\ref{sec:disc_size}. 

This means that any ejected material will either be swept up by the accreting discs or will escape the cluster, which would leave the cluster appearing to be Ôgas freeÕ.  We note that supernovae will occur at a rate that is much higher ($20-100$~times) than the time for low-mass stars to sweep out the full volume of the core. Therefore, for this model to work, the coupling between the supernovae ejecta and the dense enriched material from binary/spin-stars needs to be low, i.e., the SNe ejecta should not remove the dense ejecta.  This is supported by hydro-dynamical models (Rogers \& Pittard~2013, see also \S~\ref{sec:assumptions}).



\subsection{Mass Budget}
\label{sec:mass_budget}

As mentioned in \S~\ref{sec:intro}, models that invoke the formation of a second generation from material processed by a first generation all suffer from a ``mass budget problem", i.e. that the required mass of the first generation is significantly higher than that observed today.  The proposed solution is that the first generation was initially much more massive than observed today and that much of this mass was lost, due to the dynamicall evolution of the GC, to the field (e.g., Vesperini et al.~2010; Conroy~2012).  However, this violates observations of the fraction of stars in GCs and the field (at a given metallicity) in the Fornax Dwarf galaxy (Larsen et al.~2012 - the "external mass budget"), unless if extreme assumptions are imposed, such as all ``2nd generation" stars are still alive (i.e., the IMF of the ``2nd generation" only contained stars up to 0.8\msun) and that 100\% of the enriched material forms stars (i.e., a 100\% star-formation efficiency - D'Antona et al.~2013).

Depending on the ISM density and relative velocity of the star, a PMS star may accrete a substantial fraction of its mass over its PMS lifetime.   For example, if a 0.25 \msun\ star (with a helium mass fraction of Y=0.24) accretes 0.25 \msun\ of enriched material (Y=0.37 - dM09), then the resulting 0.5 \msun\ star will have Y$\approx0.31$, similar to that observed in NGC~2808 (e.g., Piotto et al.~2007).  Additionally, if the extreme He abundance ejecta is used from the interacting binary (Y=0.64), even more extreme values can be produced.  

\begin{figure}
\includegraphics[width=8cm]{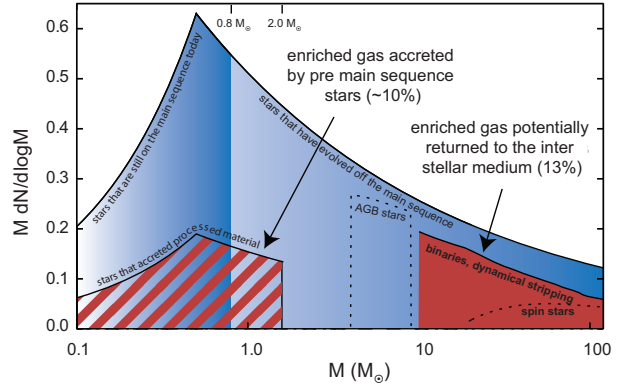}
\caption{The mass weighted initial mass function of stars in a cluster.  The amount of chemically enriched gas potentially returned to the interstellar medium from interacting high-mass binaries (13\% shown in red) is comparable to the total mass contained in low-mass stars that are observed to be enriched (19\% shown in red hashing). However, since pre-main sequence stars are already in place when the accretion begins, this is an upper limit to the amount of material needed.  If each star, on average, accretes half of its final mass from enriched material, this fraction would fall to $\sim10$\% (see \S~\ref{sec:mass_budget}). Hence, enough mass is available from the high-mass binaries to account for the mass observed in enriched stars.  The kink in the mass distribution at $0.5$~\msun\ reflects the kink in the adopted initial mass function (Kroupa~2001).}
\label{fig:mass_budget}
\end{figure}

For the calculations below, we assume that a large fraction of high mass stars are in binaries that will interact during their evolution, consistent with observations of local high mass star forming regions (Sana et al.~2012).  For a $10^6$ \msun\ cluster, approximately $1.3 \times 10^5$\msun\ of processed material from high-mass star binaries may be returned to the ISM (dM09) within 20 Myr (the lifetime of a 10 \msun\ star - Brott et al.~2011 - see Fig.~\ref{fig:mass_budget}).  If we restrict the time where discs may accrete to the first 8~Myr, the amount of material potentially available drops to $6.5 \times 10^4$\msun\ (see Fig.~\ref{fig:shedded_mass}).

For a typical massive GC like NGC 2808, approximately a third of the low mass stars show significant enrichment (e.g., Piotto et al.~2007).  Low mass stars make up 58\% of the initial mass of a stellar population (Kroupa~2001), meaning that only $\sim19$\% (i.e., 1/3) of the initial mass of the population ($\sim30$\% of the total number of stars) needs to be significantly chemically enriched.  This is already similar to the 13\% of initial cluster mass that could feasibly be processed by high mass binaries and spin stars.  However, since a PMS star is already in place at the time that accretion begins (see \S~\ref{sec:tail_end}), we can make the extreme approximation that half of its final mass will be accreted (and half comes from the starÕs initial mass), hence only $\sim10$\% of the initial cluster mass is needed. 

Additionally, since the enhanced stars are preferentially located in the core, non-enriched stars would be more easily lost due to tidal effects (e.g., Vesperini et al.~2010), further decreasing the mass fraction needed for enrichment.  Hence, we do not need to assume that all high-mass stars are in binaries, or that the clusters were initially much more massive than observed today.  As will be shown in \S~\ref{sec:abundances}, the majority of the stars in GCs have abundances that require significantly less than 50\% of their final mass to be from enriched material that has been accreted.  Hence, the numbers present here are an upper limit to the mass budget required.

In the above estimates we have assumed that low-mass stars can accrete processed material for up to 20~Myr.  If on the other hand, we limit the accretion process to the first $\sim8$~Myr (i.e., due to potentially short disc lifetimes, see \S~\ref{sec:assumptions}), the amount of material available from interacting binaries is half of that used above (see Fig.~\ref{fig:shedded_mass}).  Even in this case, we still do not violate the mass budget constraints imposed by the number ratio of the enriched/non-enriched stars or the global mass budget constraints imposed by the observations of the Fornax dwarf galaxy (Larsen et al.~2012).

\subsection{Efficiency of Accretion}
\label{sec:efficiency}

The above calculations assumed that $100$\% of the enriched material ejected from interacting binary stars is accreted onto low-mass PMS stars.  In reality, we expect that some fraction (perhaps the majority) of the enriched material is not accreted, but is lost from the cluster, either escaping due to not encountering a star with disc or being removed by SNe, ionisation or fast winds of hot stars.  As will be discussed in \S~\ref{sec:nao}, the abundances of individual elements (with the possible exception of O) do not require more than 50\% of the star's final mass to be accreted (enriched), and in many cases, is significantly less.  Hence, in more realistic cases like the one outlined below (where typically each enriched star only obtains a relatively small fraction of its final mass through ``tail-end" accretion) it is not necessary that all the enriched material is accreted, in fact it is likely that a large fraction is lost from the cluster.

The efficiency of the accretion process (i.e. the fraction of mass accreted onto the discs of young stars) depends on the cluster density (i.e., the number of stars, within a given volume, with discs that are available to sweep up the enriched material) as well as (and related to) the velocity dispersion (the faster the stars move, the larger the volume that they can sweep up).  Hence, a prediction of the proposed model is that more massive (i.e. on average denser and higher velocity dispersion) clusters should have a higher enrichment, i.e. that the stars that are able to accrete enriched material, accrete more of it.

This prediction is in accordance with observations that more massive clusters (present day masses) have a broader extension of the Na-O anti-correlation than lower mass clusters (Carretta et al.~2010).  Additionally, the extension of the horizontal branch (a signature of enrichment) is correlated with cluster mass (e.g., Gratton et al.~2010).  This also explains why lower mass open clusters do not show significant abundance spreads, as due to their significantly lower densities and velocity dispersion relative to GCs, the proposed scenario would be extremely inefficient within them.

\subsection{Chemical Abundances}
\label{sec:abundances}

\subsubsection{Nature of the polluters}
\label{sec:nature}

If interacting binaries are responsible for the polluting material, we expect significant variations between different clusters, due to stochastic effects. The yield of a binary system depends on at least three parameters: the total mass of the stars, the mass ratio and the phase in which the two stars interact.  Even in a massive globular cluster where the top end of end of the IMF is relatively well sampled, the three dimensional parameter space of binaries is not well sampled.   Therefore stochastic fluctuations from cluster to cluster are expected. 

In addition to the ejection from massive binary stars, there are multiple alternative routes to get chemically enriched material within the cluster during the first $\sim10$~Myr of a clusterÕs evolution.  Massive binaries can create spin-stars (de Mink et al.~2013) through mass transfer and stellar mergers, which can shed material (Decressin et al.~2007).  Chains of stellar collisions involving massive stars may shed material into the cluster (Glebbeek et al.~2009), which may be a way to ÒquantiseÓ the chemical enrichment and explain differences between globular clusters (e.g., Gratton et al.~2012).  This may allow material deep within the stars that is Mg depleted and Al-enriched to be liberated and shed into the cluster.  Finally, single stars that have a near/partial collision may strip the outer envelope of a star (Sills \& Glebbeek~2010). The lowest mass stars ($<0.5$~\msun) have extremely long pre-main sequence lifetimes, and may be able to continue accreting for tens of Myr.  Hence, the ejecta of AGB stars may also enrich these stars, although due to the lack of discs around these low mass stars at these ages ($>30$~Myr), accretion would be much less efficient.

\begin{figure}
\includegraphics[width=8.5cm]{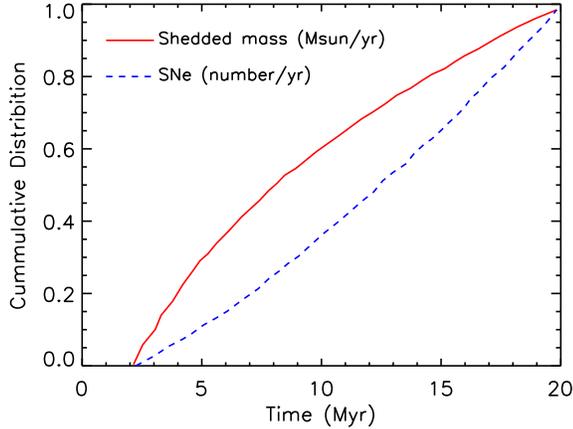}
\caption{The cumulative distribution of material ejected from interacting binaries and the cumulative number of Type II SNe, normalised between 2 and 20~Myr.  The mass lost through interacting binaries is not uniform in time, but is skewed towards young ages.  Approximately 50\% of the enriched material is shed within the first $\sim8$~Myr of the clusterÕs life.  Hence, a significant amount of material is available to be accreted by circumstellar discs on short timescales.}
\label{fig:shedded_mass}
\end{figure}

In Fig.~\ref{fig:shedded_mass} we show the cumulative distribution of the enriched ejecta (from the binary interaction model adopted in the main text) and the cumulative number of Type II SNe based on stellar evolutionary models of high mass stars (Brott et al.~2011), assuming a Salpeter~(1955) index for the initial stellar mass function.  For this we assumed the shedded mass from binaries and the Type II SNe happen after a main-sequence lifetime, although for the interacting binary case in reality some will interact before/after the end of the main sequence phase.  Additionally, we assume that the amount of ejecta for the interacting binary scales linearly with the mass of the primary. (dM09), .  Finally, we adopt main sequence lifetimes of $t_{\rm ms} \propto M/L \propto M^{1-x}$, with $x=2.0$ (although we note that the results do not depend strongly on x). 

This shows that $\sim50$\% of the shedded mass from interacting binaries (i.e. the enriched material) is lost within the first $\sim8$~Myr of a clusterÕs existence.  One of the main uncertainties in the presented model is the lifetime of the circumstellar discs (see \S~\ref{sec:assumptions}).  This shows that even if discs are truncated within the first 10~Myr years, a large fraction of the total enriched material will already have been ejected, and presumably accreted.  Fig.~\ref{fig:shedded_mass} also shows that the rate of SNe lags behind that of the interacting binaries, with a median age of $\sim12$~Myr (when normalised between 2 and 20~Myr), relative to the interacting binary, which eject half of the material within the first 8~Myr.

Many GCs show discrete main sequences that have been interpreted as being due to quantised He abundances.  As discussed in \S~\ref{sec:distribution}, this potentially may be caused by the orbits of the stars in the cluster, as only stars that enter the core may be enriched.  Additionally, some quantisation may come from the ÔstepÕ nature of the abundance patterns of the processed material from high-mass binaries (dM09), i.e., the enriched material from high-mass binaries is ejected discretely in time, with the most enriched material being ejected at the end. 

 In the proposed model, light elements (e.g., Na, O and Al) should also show the similar quantised abundance trends, as is observed in some GCs (e.g., Marino et al.~2011), although due to the low initial abundances, the accretion of a small amount of enriched material would produce relatively large abundance changes.  Hence, we expect some variations in light element abundances within GCs (i.e. not fully quantised).   Additionally, if some enriched material escapes the core (or if the orbital distribution is radially anisotropic), material may be accreted by stars in the outer parts of the cluster.  This would not affect He (since a large amount must be accreted before any effects are seen) but other elements (Na, O, Al) could be significantly affected due to the low initial abundances, hence abundance spreads are expected, not fully quantised values.

\subsubsection{Na-O anti-correlation and the mass budget imposed by observations}
\label{sec:nao}

Decressin et al.~(2007) have investigated the expected abundance trends for material ejected by spin-stars.  Additionally, dM09 computed the yields of one (typical) binary system, with stellar masses of 15 and 20~\msun, and an initial period of 12 days.  These works have shown that the enriched material from high mass stars follows the observed abundance trends.  In particular, the models are enhanced in He, N, Na, and Al, and depleted in C and O.   A caveat that is often noted is that high-mass stars do not, at least {\it a priori}, deplete Mg sufficiently.  We will address this point in detail in \S~\ref{sec:mgal}.

For the following analysis, we use the abundances from the interacting binary model presented by dM09.  This consisted of a binary with primary and secondary masses of 20 and 15~\msun, respectively, on a 12~day orbital period.  We note, however, that these simulations do not cover the full parameter space of interacting binaries (see \S~\ref{sec:nature}) and that higher mass stars or stars that interact at different phases of their evolution may produce significantly different yields than those used here. ÊWe adopt initial (i.e., pristine) abundances of  Y=0.24, [O/Fe]$ = 0.53$, [Na/Fe]$=-0.11$, [N/Fe]$=0.05$, [C/Fe]$=-0.13$ and [Al/Fe]$=0.31$ (from Piotto et al.~2007 for He, and Villanova \& Geisler~2011 for all other elements).

As a first step, we look at whether the model can reproduce the extreme ends of the abundance trends observed.  We adopt the ``extreme abundances" of the ejecta in the dM09 model.  Additionally, we assume that in these extreme cases, the star accretes half of its final mass from the enriched material.

With these assumptions we find that the extreme values potentially available with the adopted model are: Y=0.44, [O/Fe]$ = 0.26$, [Na/Fe]$=1.22$, [N/Fe]$=1.2$, [C/Fe]$=-0.4$ and [Al/Fe]$=0.62$.  In order to test whether the model can reproduce the observed values, we compare our predictions to observations of M4 (Villanova \& Geisler~2011), whose absolute values and spreads in the abundances discuss appear typical amongst GCs.  They find extreme values of [O/Fe]$ = 0.1$, [Na/Fe]$=0.57$, [N/Fe]$=0.95$, [C/Fe]$=-0.48$ and [Al/Fe]$=0.66$.

For He, we note that our extreme value is in excess of that inferred from the bluest main-sequence track in NGC~2808 (Piotto et al.~2007).  For the other elements, the model predicts the correct trends (dM09), and for Na and N, the observed extreme values are within the allowed range of the model.  For C and Al, our predictions are within 0.1~dex, hence are consistent within the observational uncertainties and model assumptions.  For O, our toy model misses by 0.3 dex, potentially not accommodating the observations.  However, as noted above, the exact predictions depend on multiple binary parameters, and so this should be further investigated in a future work.

We conclude that in addition to not violating the internal mass budget (i.e. the constraints imposed by the ratio of the enriched/unenriched populations), the proposed model also obeys the observational constraints for individual chemical abundances.  The only element not reproduced in full, is O.  Hence, no stellar IMF variations need to be invoked in the proposed scenario.

A similar scenario was investigated by Prantzos \& Charbonnel~(2006), their ``scenario II", where heavily processed (taken as the most extreme [O/Na] observed in NGC~2808) material from a ``first generation" is combined with pristine gas in the form of low mass proto-stars.  They find that the IMF from the first generation must by flatter (i.e., more high mass stars relative to a nominal Salpeter~(1955) IMF) in order to reproduce the number (and degree) of enriched stars.  

However, as discussed above, the scenario proposed here does not require IMF variations.  The main difference between our estimates and those of Prantzos \& Charbonnel~(2006) is the chemistry of the enriched material.  Whereas they took the observed limit (i.e., the most extreme [O/Na] value observed), our estimates are based on models of the ejecta of interacting binaries.  This ejecta has more extreme values of the abundances considered, so that when it is mixed with pristine material, the observed trends are reproduced.  Put another way, Prantzos \& Charbonnel~(2006) assume that the stars with the most extreme abundances come exclusively from enriched material (i.e., no pristine material is used) whereas in our scenario, the most extreme abundances are still reproduced with $50$\% pristine material.

\subsubsection{Abundance trends and Li}
\label{sec:li}

A potential caveat to high mass stars being the source of the enriched material (either as spin stars or massive close binaries), is that Li is destroyed in high-mass stars, while observations show that Li does not vary strongly within the different populations within GCs (with $\sim0.3$~dex).  We first note that in the ``early disc accretion" scenario, a low-mass star is largely in place when the accretion occurs.  Hence, in the extreme situation where a star obtains half of its final mass from accreted material (and assuming this material has no Li), the star would have a Li abundance lower by 0.3~dex.

As discussed above, the accreted material (assumed to be Li depleted) is expected to be enhanced in He, N, Na and Al, and depleted in C and O.  Hence, we expect correlations between Li and the other light elements that show variations.  Such a slight anti-correlation between between Li and Na and a correlation between Li and O has been observed in NGC~6752 (Pasquini et al.~2005).  Additionally, Monaco et al.~(2012) have studied the Li abundance of main-sequence and sub-giant branch stars in M~4 (i.e., stars not yet affected by the dredge up of evolved stars which significantly depletes Li within stars) and find that Li varies between stars (by 0.1 dex) and is anti-correlated with Na, as predicted by our model.

We conclude that while it has been claimed that the small Li variations observed in GCs argues against massive stars as the source of the chemically enriched material, this only holds if a new star is formed exclusively out of the enriched material.  In the case where only a fraction of the star's final mass comes from enriched material, the expected variation in Li is $<0.3$~dex, within the current observational uncertainties.

\subsubsection{The (anti-)correlation of Mg and Al}
\label{sec:mgal}

In addition to the well established anti-correlation between Na and O, some GCs, such as NGC 6752 and NGC 2808, show an anti-correlation between Mg and Al (Carretta et al.~2012a and Carretta et al.~2009a, respectively).  This is a potential problem for any scenario attempting to explain the multiple populations within GCs that invoke the processed material of high mass stars (such as spin-stars or interacting binaries) as the temperatures thought to be required to significantly deplete Mg and enhance Al are not reached in most high-mass stars (Decressin et al.~2007; dM09).  Although we note that stars with masses exceeding 40$\msun$\ do reach the required high temperature, although deeper within the star than would normally be ejected by winds or interacting binaries (Yusof et al.~2013).   However, such models do predict a weak Mg-Al anti-correlation, which could be made stronger by adjusting the nuclear reaction rates of Mg and Al (Decressin et al.~2007), or if the Mg isotope ratio favours $^{25}$Mg and $^{26}$Mg over $^{24}$Mg (dM09).   On the other hand, models that invoke AGB stars as the origin of the enriched material may predict such an anti-correlation, depending on the adopted parameters in the AGB evolution models (D'Ercole, D'Antona \& Vesperini~2011).  However, it may then be difficult to explain why most clusters do not show such a trend, and only some do.

A closer look at the available data reveals that the Mg-Al anti-correlation (in particular, stars extremely depleted in Mg) appears to be the exception, rather than the rule.  In a compilation of 17 clusters, only three were found to have statistically significant variations in Mg (Carretta et al.~2009a - less than 3\% of the sample of stars showed significant Mg depletion), and a number of other recent studies that have searched for such a trend have not found a Mg-Al anti-correlation nor evidence for significant Mg variations within individual clusters (e.g., M4 - Villanova \& Geisler~2011; 47~Tuc - Gratton et al.~2013; M22 - Marino, Milone \& Lind~2013; NGC~3201 - Mu\~noz, Geisler, \& Villanova 2013, M75 - Kacharov, Koch, \& McWilliam~2013).   Another cluster, NGC 1851, shows evidence for a small Mg-Al anti-correlation, although it does not have any stars strongly depleted in Mg (Carretta et al.~2012b).

Based on the available data, it appears that the majority of GCs do not show a strong Mg-Al anti-correlation, and only a small subset of GCs contain stars depleted in Mg.   Hence, the model presented here fits the observations of Mg and Al for the vast majority of GCs observed to date, and potentially may explain the rarity of stars strongly depleted in Mg.

Large Mg spreads (being anti-correlated with Al content) may be expected in the proposed model if the polluting material is due to the collisions of high mass stars (which can liberate significant amounts of Al-rich, Mg-depleted material from deep inside the star - Sills \& Glebbeek~2010).  Also, if the polluting material has been enriched by very massive stars, a similar effect is expected (Yusof et al.~2013; Denissenkov \& Hartwick~2013).

\subsection{Fe Spreads in Clusters and Constraints from the Solar System}
\label{sec:fe}

The ``tail-end" accretion scenario was originally developed to explain peculiar abundances observed in the Orion Association (within the same sub-group) between young stars (Throop \& Bally~2008).  Due to the observed abundance trends, the authors invoked Type~II supernovae and the outflows of massive stars as the source of enrichment.  If SNe, are indeed able to enrich discs then we may expect to find some evidence for spreads in heavy elements, such as Fe, that are uniquely produced in SNe.

Observations of $^{60}$Fe in meteorites suggest that the early solar system (when the sun still had a proto-planetary disc) was enriched by a nearby SNe (see Adams~2010 for a recent review).  This evidence for SNe enrichment in the early solar system is one of the primary pieces of evidence used to reconstruct the birth environment of the solar system (e.g., Throop \& Bally~2008; Adams~2010; Dukes \& Krumholz~2012).

If discs are able to accrete some material from SNe ejecta (as appears to be required by the early solar system), then we may expect that clusters with higher densities, i.e., the covering fraction of discs seen by the SNe ejecta is higher, may show some enhancement in heavy elements, such as Fe.  Here, the velocity dispersion of the cluster is not likely to be relevant, due to the high velocity of the SNe ejecta.

Carretta et al.~(2009b) have shown that the intrinsic spread in Fe abundances between stars in GCs is correlated with cluster mass.  This is qualitatively consistent with the scenario outlined above, as higher mass clusters tend to be denser clusters, although their initial densities may have been significantly altered by a Hubble time of dynamical evolution (Giersz \& Heggie~2009; Gieles, Heggie, \& Zhao~2011).  Simulations probing the efficiency of the retention of SNe ejecta by circumstellar discs is required before any quantitative prediction can be made.  

\section{Assumptions and Potential Caveats}
\label{sec:assumptions}

There are a number of assumptions and potential caveats for the early disc accretion model, and here we list and discuss some of the primary ones.

\subsection{The lifetime and size of circumstellar discs}

As discussed in \S~\ref{sec:model}, Bondi-Hoyle accretion is not efficient enough in young GCs for the proposed model to work.  Hence, it is necessary for the low mass PMS stars to host circumstellar discs in order to sweep up the processed material.  If stars lose their discs too rapidly, then the proposal mechanism/scenario will not work.

\subsubsection{The lifetime of circumstellar discs}

While disc lifetimes are seen to be 5-15 Myr for low mass stars in the solar neighbourhood (e.g., Haisch, Lada \& Lada~2001; Bell et al.~2013) there are theoretical reasons to believe that the disc lifetime may have been different in GCs.  Firstly, if there is enough ambient material to accrete onto the disc from the surroundings, the disc may maintain (or even increase) its mass (Moeckel \& Throop~2009).  The density within the cores of globular clusters is much higher than in nearby star-forming regions and there are also many more massive stars available in young GCs to provide enriched material.  Hence the accretion rate onto discs from the surrounding material is expected to be orders of magnitude higher than in local star-forming regions.  This accretion onto the disc may increase the accretion rate onto the host star as well as increase the disc lifetime (see \S~\ref{sec:nature}). 

Secondly, disc lifetimes may depend on the metallicity of the system, with suggestions that in low metallicity environments, where GCs were born, the disc lifetimes may have been significantly longer than at present in the galaxy.  This is potentially due to 1) viscous accretion being less efficient at low metallicities (Durisen et al.~2007)  2) lower metallicity environments having less dust so grain growth is suppressed (Dullemond \& Dominik~2005) limiting the effects of photo-evaporation, and 3) planet formation, which appears to increase towards higher metallicity, has been suggested to be a main disperser of circumstellar discs (e.g., Armitage \& Hansen~1999). Additionally, there has been tentative suggestions, based on observations of pre-main sequence stars in various environments, that disc lifetimes are much longer in low-metallicity environments (Spezzi et al.~2012).  

However, it is also possible that disc destruction is more effective in dense environments like globular clusters.  The presence of significant amounts of nearby high mass stars, may cause the discs to photo-evaporate faster than in locally observed regions (e.g., Adams et al.~2004).  Additionally, the destruction of discs by close encounters is more efficient in denser regions (although we note it is less efficient as the velocity dispersion increases), which we address below.

\subsubsection{The size of circumstellar discs}
\label{sec:disc_size}

Discs around young stars are observed to have radii between a few tens of AU and thousands of AU (Lada et al.~2000)  The main truncation mechanism for discs in a dense cluster environment is likely to be the close passage of another star in the cluster.  This has been studied by a number of authors, generally for relatively low mass (low number of stars) clusters (de Juan Ovelar et al.~2012; Scally \& Clarke~2001; Olczak, Pfalzner, \& Spurzem~2006).  The truncation radius is a function of the pericentre distance between the host star and the perturber star, the starsÕ relative mass ratio, and the velocity difference between the two stars.  As an example, we take the recent analytical estimates of disc truncation in young star forming regions in the solar neighbourhood (de Juan Ovelar~2012), and apply a correction factor of 10-20 (i.e., we assume that the truncation radius scales linearly with the velocity dispersion, although see below) to account for the much higher velocities expected within young globular clusters (e.g., Bastian et al.~2006).

For a million solar mass cluster, with a core radius of 2pc, the average stellar surface density within the core (containing $\sim12$\% of the stars at any given time) is $\sim2 \times 10^4$ stars pc$^{-2}$.  In this example, after 1~Myr the truncation radius for an average star is expected to be $\sim2000$~AU.  This decreases by a factor of $\sim5$ every 3~Myr (de Juan Ovelar et al.~2012), resulting in a radius of $\sim80$~AU in 7~Myr.  

However, this may overestimate the truncation for three reasons.  i) The simulations of dynamical disc truncation carried out to date were of low-N, low velocity dispersion clusters.  This means that the duration of a disc/star interaction is a significant fraction of the disc orbital timescale, maximising the effect of disruption.  In high velocity dispersion clusters, the interaction duration is significantly less than an orbital timescale, leading to a reduced disruption/truncation effect.  In reality, in high-$\sigma$ clusters, the dynamical truncation of discs may not begin until the average encounter distance is less than twice the disc radii, i.e. the discs themselves begin to touch during the encounter.  ii) Even in those encounters, the interaction may lead, instead of a pure truncation and stripping the material, to mass from one disc being transferred to the other disc (Clarke \& Pringle~1993).  iii) The discs in the core of a massive GC are likely to be accreting significant amounts of material (i.e. the processed ejecta of massive binary stars) which add mass to the accreting disc (Moeckel \& Throop~2009). 

We note that denser (core) clusters will likely have more heavily truncated discs.  However, this effect is counter-acted, so some extent, by the fact that (for a given cluster mass) a smaller volume will need to be swept out by the PMS disc.

\subsubsection{The effect of ram pressure stripping and supernovae}
\label{sec:ram}

In order to estimate the effect of ram pressure stripping of the discs, we take as an example a disc of mass 0.05~\msun\ around a 0.5~\msun\ star.  A `standard' disc has a surface density profile $\Sigma( r ) \propto r^{-3/2}$ (e.g. a standard minimum mass solar nebula - Weidenschilling 1977), and let us assume that $\Sigma(1 AU) = 3.4 \times 10^4$ kg/m$^{2}$ (which contains the total disc mass within 100 AU).  We will assume an extreme case of the disc traveling face-on at 30~km\,s$^{-1}$ through a medium of density $6 \times 10^{-19}$ g/cm$^{3}$  (10~\msun\ of ejecta in $10^{-3}$~pc$^3$), similar to that expected for the ejecta from interacting binaries (e.g., Smith et al.~2007).  

The radius at which the ram pressure of the inter-cluster medium is able to strip the disc is $>100$~AU (Moeckel \& Throop~2009).  The addition of momentum to the disc is important if the disc can accrete of order its own mass in of order an orbital time of the disc.  The time for a $100$~AU disc to accrete $0.05$~\msun\ is $0.25$~Myr in our extreme case -- many times the outer orbital timescale.  It should be noted that the accreted material will be of lower angular momentum than the disc material and cause contraction of the disc and enhanced accretion (Lin \& Pringle 1990; Moeckel \& Throop~2009).

Similar calculations show that the ram pressure of material from a supernova explosion will not significantly strip a disc around a low-mass star unless the star is extremely close (much less than 0.1\,pc) to the supernova (Chevalier~2000). 

Therefore we do not expect discs around low-mass stars to be truncated by the effect of the external interstellar medium, rather the limitation on discs sizes will be set by encounters with other stars/discs.

Additionally, the enriched material from interacting binaries will be preferentially ejected in a disc.  As it expands, the material is expected to be quite dense (see above), which, as in the case of circumstellar discs, is not expected to be strongly influenced by nearby supernovae, due to the low coupling between the SNe energy and cooler, dense material (Rogers \& Pittard~2013).

In summary, the detailed processes of disc accretion (sweeping of material), how the disc responds to its environment (e.g., dynamical interactions, photo-evaporation, SNe), and the fate of the accreted material (i.e. if it will end up on the host star) need to be further investigated with numerical simulations to determine whether the proposed model can fully explain the observed chemical anomalies in GCs.

\subsection{Initial Conditions}
\label{sec:initial}

We have assumed that the cluster is gas free and mass segregated from an early age ($<3$~Myr).  These conditions are taken from known properties of young massive clusters (with masses up to $\sim10^5$\msun) in the Galaxy (see the recent review in Longmore et al.~2013).  Many/most young clusters appear to have their massive stars concentrated towards to the centre (e.g., de Grijs \& Parmentier~2007; Portegies Zwart et al.~2010) although quantifying the exact amount has been difficult due to observational biases (e.g., Ascenso et al.~2009).  

In the estimates for the gas clearing time within the cluster core, we adopted a core radius of $1$~pc (\S~\ref{sec:accretion_rate}).  This is consistent with observations of YMCs in the Galaxy with masses up to $10^5~\msun$ (Portegies Zwart et al.~2010) and also with extragalactic YMCs with masses between $5 \times 10^5$ and $1.5 \times 10^6~\msun$ (e.g., Larsen et al.~2001, 2008, 2011; McCrady, Graham \& Vacca~2005).  This is also consistent with simulations of the evolution of massive globular clusters, which were found to have been much denser in the past than they currently are, e.g., R$_{core, initial} < 0.3$~pc for 47~Tuc, M4 and NGC~6397 (Giersz \& Heggie~2009, 2011). 

The mechanism that has caused these young clusters to be gas free from a young age is still under debate.  Potentially the gas and stars have become kinematically decoupled during the cluster forming debate, resulting in an extremely high {\em local} star-formation efficiency (e.g., Moeckel \& Clarke~2011; Kruijssen et al.~2012).  Alternatively, it has been suggested that feedback from high mass stars (or low mass stellar outflows) has caused the gas to be rapidly removed, although this predicts that many young clusters should be out of dynamical equilibrium (i.e.g, expanding) which is not supported by observations (c.f., Longmore et al.~2013).

If it is a locally high star-formation efficiency that is causing the clusters to be gas poor at a young age, then this will have no impact on the discs around young low-mass stars or on the ejecta of interacting binaries/spin-stars.  If the gas is cleared by feedback, then potentially the same effect that clears the cluster of the natal gas may effect the discs/ejecta in a similar manner, if the natal gas (at the time of gas dispersal) had similar density as the interacting binary/spin-star material.

\subsection{The Relation Between Globular Clusters and Dwarf Nuclei}
\label{sec:fe_spreads}

In \S~\ref{sec:fe} we discussed a potential mechanism that would allow clusters to self-enrich in elements produced in SNe (i.e., the retention or SNe ejecta by circumstellar discs which has been invoked to explain the meteoritic abundances in the solar system).  This would allow clusters to display spreads in elements that are (nearly) solely produced in SNe, such as Fe.  An alternative possibility is that clusters that display large Fe spreads are the captured nuclei of dwarf galaxies.


If the latter scenario is true, then those GCs that spent a significant fraction of their existence at the centres of deep gravitational potential wells, and as such, could accrete gas (and stars) from their host galaxies (e.g., Kruijssen \& Cooper~2012).  Nuclear clusters in low-mass galaxies in the local Universe show clear evidence for multiple (or extended) star forming events, so that their stellar mass builds up over time (e.g., Rossa et al.~2006; Seth et al.~2008).    While these clusters probably underwent similar self-enrichment as discussed here, they also had multiple star-forming events (fed by pristine gas from the host galaxy as well as that enriched by massive and intermediate mass stars, i.e., AGB stars), hence are expected to have potentially different abundance patterns for certain elements.  

Hence, at the moment it is unclear whether the ``early disc accretion" scenario applies to clusters with significant Fe spreads, or whether these clusters are from an extragalactic origin.

\subsection{Relation to the Intermediate Age Clusters in the LMC/SMC}
\label{sec:intermediate}

A number of recent studies have shown that the main-sequence turn-off in intermediate age ($1-2$~Gyr) LMC/SMC clusters, with present day masses between $10^4 - 2\times10^5$\msun, is broader than would be expected from a single aged population (e.g., Mackey \& Broby Nielson~2007).  This has been interpreted as being due to an extended, 100-500~Myr, star-formation history within the clusters (e.g., Mackey et al.~2008; Milone et al.~2009; Goudfrooij et al.~2011).  Theoretical studies have attempted to link this phenomenon with the chemical anomalies observed in GCs (e.g., Conroy \& Spergel~2011; Keller, Mackey, \& Da Costa~2011).  If the extended main-sequences and dual red-clump observed in some clusters (e.g., Girardi, Rubele, \& Kerber~2009) is due to extended (or multiple burst) star-formation histories, then this would show that clusters can retain the material needed for additional star-formation episodes and that the chemical anomalies in GCs may be due to secondary or extended star-formation events.

A prediction of the above interpretations of significant age spreads is that massive clusters with ages $<500$~Myr should be currently forming stars.  Bastian \& Silva-Villa~(2013) investigated two massive clusters in the LMC, NGC~1856 and NGC~1866 (both with $\sim10^5$~\msun, and ages of $280$~Myr and $180$~Myr, respectively) with HST resolved stellar photometry, and found that both were consistent with a single burst of star-formation at the time of formation.  The authors derive an upper limit to any potential age spread within the cluster of $35$~Myr. 

Additionally, Bastian et al.~(2013) have compiled a catalogue of 129 young massive clusters with ages between 10 and 1000~Myr and masses between $10^4$ and $10^8$\msun, with available spectroscopy and resolved photometry to search for signs of on-going star-formation within them (i.e., $H\beta$ and/or O[{\sc iii}]$\lambda \lambda$4959,5007 emission - see also Peacock et al.~2013).  No clusters were found with any evidence of on-going star-formation.  Based on stellar IMF sampling (i.e., taking into account the mass of the secondary population) and the observational limit where emission  lines could be detected, Bastian et al.~(2013) showed that at least half of their compiled catalogue (with ages less than $\sim200$~Myr) should show clear emission lines associated with on-going star-formation, if the clusters had extended star formation histories similar to that reported for the LMC/SMC intermediate age clusters.  

The authors conclude that either the stellar IMF within the secondary population (i.e., that currently forming) is radically different from that observed locally (i.e. no stars with masses above $\sim15$\msun\ can form within them) or that even massive clusters do not have extended star formation histories.  In the latter case, this calls into question the interpretation of the extended main sequence turn-off in intermediate age LMC/SMC clusters is due to extended star formation events.  The observed spread may instead be caused by the enrichment mechanism proposed here, or potentially through stellar rotation (Bastian \& de Mink~2009; Yang, Bi, \& Meng~2013; although also see Girardi, Eggenberger, \& Miglio~ 2011). 

\section{Predictions of the Model}
\label{sec:predictions}

Some of the predictions and tests of the proposed model, such as the chemical abundance trends, have been previously discussed (dM09).  Here, we discuss further predictions of the model, some are unique to the proposed scenario, while others are similar to predictions from previously proposed scenarios. 

A) The ``early disc accretion" model naturally explains why Na-rich, O-poor stars are not observed in the halo of the galaxy (e.g., Gratton et al.~2012) as stars with this abundance pattern would preferentially exist in the cores of massive clusters.  As clusters move in an external tidal field, they preferentially lose stars from their outer parts (e.g., Vesperini et al.~2010).  However, all the main scenarios make the same prediction.


B) A prediction of the model is that massive clusters forming today should show a similar effect.  However, due to their high abundances (near solar) small changes in the Na, O, or Al abundance would not be readily detectable, since abundances are measured on a logarithmic scale.  However, some of the low-mass stars may be He enriched, hence young massive clusters with ages $>20$~Myr (e.g., Glimpse-C01 with an age of a few-hundred Myr - Davies et al.~2011) may be used to test this scenario.  At solar metallicity, He enriched stars are offset from the nominal (standard He abundance) main sequence in optical colour-magnitude diagrams by a similar amount as that seen in a fraction of old Globular Clusters (Dotter et al. 2008).  This should be visible in deep colour magnitude diagrams of young ($<1$~Gyr) massive clusters (if the stars within the cluster kept their discs long enough to be enriched). Potentially, they should show the Na-O (anti-)correlation as well, although this effect may be drastically reduced in magnitude due to the much higher (a factor of $\sim10-100$) metallicity in YMCs today compared to GCs\footnote{A potential caveat to this prediction is that if disc lifetimes are metallicity dependent, this may not be happening in young massive clusters in spiral or merging galaxies.}.

C) If a cluster is not fully relaxed (i.e., it is younger than a relaxation time), the stars within the cluster will retain some memory of their initial orbits.  Since our model requires the concentration of high mass stars in the cluster centre, low mass stars with orbits that pass through the cluster core often (or spend a significant fraction of their orbit within the core) should be more highly enriched.  Even if a cluster is older than its relaxation time, stars in the outer parts of the cluster may retain a memory of their initial orbits.  Hence, we predict that enriched stars in the outer parts of clusters should be on preferentially radially anisotropic orbits and that stars distant from the centre that are on preferentially circular orbits should be unenriched.  There are suggestions that the more enriched stars are preferentially on radial orbits in 47~Tuc (Richer et al.~2013).  Studies with GAIA, focussing on the outer parts of the cluster, may be able to place constraints on this model, i.e. stars with orbits that do not enter the core should not show enrichment.

D) The proposed mechanism for the accretion of the enriched material is expected to be more efficient at higher cluster masses and densities.  In our model, we have assumed that all of the processed and ejected material is entrained by discs around low-mass stars.  More realistically, we expect some of the material to leak out of the cluster and become lost from the system.  In higher mass GCs, the velocity dispersion is expected to be higher (on average) which increases the amount of volume each star and sweep up.  Additionally, if a GC has a higher density, then the volume necessary to be swept out by a single star before it has accreted a significant amount of enriched material is lower.  This prediction (which is not unique among theories for the origin of multiple populations in GCs) appears to be confirmed by observations which show a correlation between the extent of the horizontal branch (a measure of enrichment) and the absolute visual magnitude of the cluster (Carretta et al.~2010), with brighter (more massive) cluster showing signs of more enrichment.

We note that in the proposed model, there is not a set density or mass for which the model operates.  Such an effect should be happening in all clusters, however the efficiency of the process is strongly dependent on the central stellar density and the velocity dispersion, with higher values of both leading to more accretion (although high central densities may also lead to a stronger disc truncation).  Low velocity dispersion clusters, such as open clusters, are expected to have some enrichment in their low-mass stars, however it is expected to be much weaker than in high mass GCs or Galactic/extragalactic young massive clusters. Additionally, the model requires the presence of significant numbers of high-mass stars in order to provide the enriched material.

\section{Conclusions}
\label{sec:conclusions}

We have presented a scenario to explain the observed abundance anomalies observed in globular clusters.  The model invokes ``tail-end" accretion of processed (chemically enriched) material shed from interacting binaries and/or rotating mass stars onto the circumstellar discs of low-mass pre-main sequence stars.  The model does not invoke multiple generations of star formation.

The proposed model potentially solves three outstanding problems with previous scenarios.  First, it solves the ``internal mass budget problem", i.e. that the enriched population makes up a significant fraction of the total number of stars within a GC.  Secondly, the solution to the ``internal mass budget problem" does not require GCs to have been significantly more massive at birth, relative to their current mass.  Hence it does not violate observations of the fraction of metal poor stars in GCs and the field in the Fornax dwarf galaxy (Larsen et al.~2012), i.e., it satisfies the ``external mass budget".  Finally, the proposed model conforms to known properties of young globular clusters (i.e., appearing gas free from a young age, no significant age spreads or secondary star formation events).


Additionally, comparison between the observed vs. predicted abundances (spreads and extreme values) shows overall good agreement, with the potential exceptions of O, and Mg in a minority of clusters.  These discrepancies may be solved if higher mass binaries were considered, or if stellar mergers/stripping plays an important role in providing the enriched gas.  Further investigation of the yields of interacting binary stars with different initial masses and separations is needed before definitive conclusions can be reached regarding the detailed chemical predictions of interacting binaries.  We have used the yields of one sample interacting binary calculation (dM09) of a 15 and 20\msun\ binary system, with an initial period of 12 days.

Potential caveats to the model were discussed. The main assumption is that discs around low-mass PMS stars survive for $5-10$~Myr.  While this conforms to observations of clusters in the solar neighbourhood (e.g., Bell et al.~2013), it is possible that the globular cluster environment is more hostile to disc survival.  Additionally, the model assumes that the discs and enriched ejecta are not strongly affected by SNe within the cluster. These caveats were discussed in detail in \S~\ref{sec:assumptions}.  Further work regarding the lifetime of discs in low-metallicity, massive GCs is required to test these assumptions.

We also discussed predictions of the model.  If the same mechanism is in operation in young globular clusters today, then some of the young massive clusters should show He enhancements (seen as spreads or separate sequences in the main-sequence with photometric studies).  Due to the higher average metallicity of young massive clusters, relative to the ancient GCs, young clusters are not expected to show significant spreads in the abundances of elements such as Na, O, Al, Fe, etc. (unless the yields increase significantly with increasing metallicity).

In addition to testing the potential caveats and predictions of the proposed model, significant progress can also be made on the observational side.  With the large GC samples now available, it should be possible to quantify the fraction of clusters that show He enrichment (and to what degree), which chemical (anti-)correlations are common to all clusters (e.g., Na-O) and which are only exhibited in a subset of GCs (e.g., Mg-Al), and what is the absolute abundance spread observed in each cluster for a large number of elements.  Once such a list is tabulated, it should be possible to isolate extreme cases (e.g., $\omega$-Cen) and look for common trends.  Until then, it is difficult to test the models directly, as it is uncertain if a given property of a cluster is unique, or a general feature of all/most GCs.

\section*{Acknowledgments}
We thank Nathan Mayne, Jay Strader, Diederik Kruijssen, Nick Moeckel, and Phil Armitage for insightful discussions.   NB and MG are partially funded by University Research Fellowships from the Royal Society.  S.d.M. acknowledges support by NASA through Hubble Fellowship grant HST-HF-51270.01-A awarded by the Space Telescope Science Institute, which is operated by the Association of Universities for Research in Astronomy, Inc., for NASA, under contract NAS5-26555 and the Einstein Fellowship program through grant PF3-140105  awarded by the Chandra X-ray Center, which is operated by the Smithsonian Astrophysical Observatory for NASA under the contract NAS8-03060. We thank the Aspen Center for Physics and the NSF Grant \#1066293 for hospitality during the conception and writing of this paper.


\begin{thebibliography}{99}

\bibitem[Adams et al.(2004)]{2004ApJ...611..360A} Adams, F.~C., Hollenbach, 
D., Laughlin, G., \& Gorti, U.\ 2004, ApJ, 611, 360 

\bibitem[Adams(2010)]{2010ARA&A..48...47A} Adams, F.~C.\ 2010, ARAA, 48, 47 

\bibitem[Allison et al.(2009)]{2009ApJ...700L..99A} Allison, R.~J., 
Goodwin, S.~P., Parker, R.~J., et al.\ 2009, ApJL, 700, L99 

\bibitem[Armitage 
\& Hansen(1999)]{1999Natur.402..633A} Armitage, P.~J., \& Hansen, B.~M.~S.\ 1999, Nature, 402, 633 

\bibitem[Ascenso et 
al.(2009)]{2009A&A...495..147A} Ascenso, J., Alves, J., \& Lago, M.~T.~V.~T.\ 2009, A\&A, 495, 147 


\bibitem[Baraffe et 
al.(2002)]{2002A&A...382..563B} Baraffe, I., Chabrier, G., Allard, F., \& Hauschildt, P.~H.\ 2002, A\&A, 382, 563 


\bibitem[Bastian et 
al.(2006)]{2006A&A...448..881B} Bastian, N., Saglia, R.~P., Goudfrooij, P., et al. 2006, A\&A, 448, 881 

\bibitem[Bastian 
\& de Mink(2009)]{2009MNRAS.398L..11B} Bastian, N., \& de Mink, S.~E. 2009, MNRAS, 398, L11 




\bibitem[Bastian 
\& Silva-Villa(2013)]{2013MNRAS.431L.122B} Bastian, N., \& Silva-Villa, E. 2013, MNRAS, 431, L122 

\bibitem[Bastian et al.(2013)]{temp} Bastian, N., Cabrera-Ziri, I., Davies, B., \& Larsen, S.S.  2013, MNRAS, submitted

\bibitem[Bell et al.(2013)]{2013MNRAS.tmp.1742B} Bell, C.~P.~M., Naylor, 
T., Mayne, N.~J., Jeffries, R.~D., \& Littlefair, S.~P.\ 2013, MNRAS, in press (arXiv:1306.3237)


\bibitem[Bondi 
\& Hoyle(1944)]{1944MNRAS.104..273B} Bondi, H., \& Hoyle, F.\ 1944, MNRAS, 104, 273 

\bibitem[Brott et 
al.(2011)]{2011A&A...530A.115B} Brott, I., de Mink, S.~E., Cantiello, M., et al.\ 2011, A\&A, 530, A115 






\bibitem[Campbell et al.(2010)]{2010MNRAS.405..421C} Campbell, M.~A., 
Evans, C.~J., Mackey, A.~D., et al.\ 2010, MNRAS, 405, 421 

\bibitem[Carretta et 
al.(2009)]{2009A&A...505..139C} Carretta, E., Bragaglia, A., Gratton, R., \& Lucatello, S.\ 2009a, A\&A, 505, 139 

\bibitem[Carretta et 
al.(2009)]{2009A&A...508..695C} Carretta, E., Bragaglia, A., Gratton, R., D'Orazi, V., \& Lucatello, S.\ 2009b, A\&A, 508, 695 

\bibitem[Carretta et 
al.(2010)]{2010A&A...516A..55C} Carretta, E., Bragaglia, A., Gratton, R.~G., et al.\ 2010, A\&A, 516, A55 

\bibitem[Carretta et al.(2012)]{2012ApJ...750L..14C} Carretta, E., 
Bragaglia, A., Gratton, R.~G., Lucatello, S., 
\& D'Orazi, V.\ 2012a, ApJL, 750, L14 

\bibitem[Carretta et 
al.(2012)]{2012A&A...543A.117C} Carretta, E., D'Orazi D'Orazi, V., Gratton, R.~G., \& Lucatello, S.\ 2012b, A\&A, 543, A117 

\bibitem[Carretta et al.(2013)]{2013arXiv1307.4085C} Carretta, E., 
Bragaglia, A., Gratton, R.~G., et al.\ 2013, A\&A, in press (arXiv:1307.4085)



\bibitem[Chevalier(2000)]{2000ApJ...538L.151C} Chevalier, R.~A.\ 2000, 
ApJL, 538, L151 

\bibitem[Clarke 
\& Pringle(1993)]{1993MNRAS.261..190C} Clarke, C.~J., \& Pringle, J.~E.\ 1993, MNRAS, 261, 190 


\bibitem[Cohen et al.(2002)]{2002AJ....123.2525C} Cohen, J.~G., Briley, 
M.~M., \& Stetson, P.~B.\ 2002, AJ, 123, 2525 

\bibitem[Conroy 
\& Spergel(2011)]{2011ApJ...726...36C} Conroy, C., \& Spergel, D.~N. 2011, ApJ, 726, 36 

\bibitem[Conroy(2012)]{2012ApJ...758...21C} Conroy, C. 2012, ApJ, 758, 21 

\bibitem[Crowther et al.(2010)]{2010MNRAS.408..731C} Crowther, P.~A., 
Schnurr, O., Hirschi, R., et al. 2010, MNRAS, 408, 731 


\bibitem[Dantona et al.(1983)]{1983MmSAI..54..173D} D'Antona, F., Gratton, 
R., \& Chieffi, A.\ 1983, Societˆ Astronomica Italiana, Memorie, 54, 173 

\bibitem[D'Antona et al.(2013)]{2013arXiv1306.3351D} D'Antona, F., Caloi, 
V., D'Ercole, A., et al.\ 2013, MNRAS, in press (arXiv:1306.3351) 

\bibitem[Davies et al.(2011)]{2011MNRAS.411.1386D} Davies, B., Bastian, N., 
Gieles, M., et al.\ 2011, MNRAS, 411, 1386 

\bibitem[de Grijs 
\& Parmentier(2007)]{2007ChJAA...7..155D} de Grijs, R., \& Parmentier, G.\ 2007, ChJAA, 7, 155 

\bibitem[Decressin et 
al.(2007)]{2007A&A...464.1029D} Decressin, T., Meynet, G., Charbonnel, C., Prantzos, N., \& Ekstr{\"o}m, S. 2007, A\&A, 464, 1029 

\bibitem[Decressin et 
al.(2009)]{2009A&A...505..727D} Decressin, T., Charbonnel, C., Siess, L., et al.\ 2009, A\&A, 505, 727 


\bibitem[de Juan Ovelar et 
al.(2012)]{2012A&A...546L...1D} de Juan Ovelar, M., Kruijssen, J.~M.~D., Bressert, E., et al.\ 2012, A\&A, 546, L1 

\bibitem[de Mink et 
al.(2009)]{2009A&A...507L...1D} de Mink, S.~E., Pols, O.~R., Langer, N., \& Izzard, R.~G. 2009, A\&A, 507, L1 (dM09)

\bibitem[de Mink et al.(2013)]{2013ApJ...764..166D} de Mink, S.~E., Langer, 
N., Izzard, R.~G., Sana, H., \& de Koter, A.\ 2013, ApJ, 764, 166 

\bibitem[Denissenkov 
\& Hartwick(2013)]{2013arXiv1305.5975D} Denissenkov, P.~A., \& Hartwick, F.~D.~A.\ 2013, ApJ, submitted (arXiv:1305.5975)


\bibitem[D'Ercole et al.(2008)]{2008MNRAS.391..825D} D'Ercole, A., 
Vesperini, E., D'Antona, F., McMillan, S.~L.~W., 
\& Recchi, S. 2008, MNRAS, 391, 825 

\bibitem[D'Ercole et al.(2011)]{2011MNRAS.415.1304D} D'Ercole, A., 
D'Antona, F., \& Vesperini, E.\ 2011, MNRAS, 415, 1304 



\bibitem[Dotter et al.(2008)]{2008ApJS..178...89D} Dotter, A., Chaboyer, 
B., Jevremovi{\'c}, D., et al.\ 2008, ApJS, 178, 89 

\bibitem[Dukes 
\& Krumholz(2012)]{2012ApJ...754...56D} Dukes, D., \& Krumholz, M.~R.\ 2012, ApJ, 754, 56 


\bibitem[Dullemond 
\& Dominik(2005)]{2005A&A...434..971D} Dullemond, C.~P., \& Dominik, C.\ 2005, A\&A, 434, 971 


\bibitem[Durisen et al.(2007)]{2007prpl.conf..607D} Durisen, R.~H., Boss, 
A.~P., Mayer, L., et al.\ 2007, Protostars and Planets V, 607 


\bibitem[Elson et al.(1987)]{1987ApJ...323...54E} Elson, R.~A.~W., Fall, 
S.~M., \& Freeman, K.~C.\ 1987, ApJ, 323, 54 

\bibitem[Fall 
\& Zhang(2001)]{2001ApJ...561..751F} Fall, S.~M., \& Zhang, Q.\ 2001, ApJ, 561, 751 

\bibitem[Gieles et al.(2011)]{2011MNRAS.413.2509G} Gieles, M., Heggie, 
D.~C., \& Zhao, H.\ 2011, MNRAS, 413, 2509 

\bibitem[Giersz 
\& Heggie(2009)]{2009MNRAS.395.1173G} Giersz, M., \& Heggie, D.~C.\ 2009, MNRAS, 395, 1173 

\bibitem[Giersz 
\& Heggie(2011)]{2011MNRAS.410.2698G} Giersz, M., \& Heggie, D.~C.\ 2011, MNRAS, 410, 2698 


\bibitem[Girardi et al.(2009)]{2009MNRAS.394L..74G} Girardi, L., Rubele, 
S., \& Kerber, L. 2009, MNRAS, 394, L74 

\bibitem[Girardi et al.(2011)]{2011MNRAS.412L.103G} Girardi, L., 
Eggenberger, P., \& Miglio, A. 2011, MNRAS, 412, L103 


\bibitem[Glebbeek et 
al.(2009)]{2009A&A...497..255G} Glebbeek, E., Gaburov, E., de Mink, S.~E., Pols, O.~R., \& Portegies Zwart, S.~F.\ 2009, A\&A, 497, 255 



\bibitem[Goudfrooij et al.(2011)]{2011ApJ...737....3G} Goudfrooij, P., 
Puzia, T.~H., Kozhurina-Platais, V., \& Chandar, R. 2011, ApJ, 737, 3 


\bibitem[Gratton et 
al.(2010)]{2010A&A...517A..81G} Gratton, R.~G., Carretta, E., Bragaglia, A., Lucatello, S., \& D'Orazi, V.\ 2010, A\&A, 517, A81 

\bibitem[Gratton et 
al.(2012)]{2012A&ARv..20...50G} Gratton, R.~G., Carretta, E., \& Bragaglia, A. 2012, A\&ARv, 20, 50 

\bibitem[Gratton et 
al.(2013)]{2013A&A...549A..41G} Gratton, R.~G., Lucatello, S., Sollima, A., et al.\ 2013, A\&A, 549, A41 

\bibitem[Haisch et al.(2001)]{2001ApJ...553L.153H} Haisch, K.~E., Jr., 
Lada, E.~A., \& Lada, C.~J.\ 2001, ApJL, 553, L153 

\bibitem[Henon(1959)]{1959AnAp...22..126H} Henon, M.\ 1959, Annales 
d'Astrophysique, 22, 126 

\bibitem[Holtzman et al.(1992)]{1992AJ....103..691H} Holtzman, J.~A., 
Faber, S.~M., Shaya, E.~J., et al.\ 1992, AJ, 103, 691 


\bibitem[Kacharov et 
al.(2013)]{2013A&A...554A..81K} Kacharov, N., Koch, A., \& McWilliam, A.\ 2013, A\&A, 554, A81 


\bibitem[Keller et al.(2011)]{2011ApJ...731...22K} Keller, S.~C., Mackey, 
A.~D., \& Da Costa, G.~S. 2011, ApJ, 731, 22 

\bibitem[Kroupa(2001)]{2001MNRAS.322..231K} Kroupa, P.\ 2001, MNRAS, 322, 
231 

\bibitem[Kruijssen 
\& Cooper(2012)]{2012MNRAS.420..340K} Kruijssen, J.~M.~D., \& Cooper, A.~P.\ 2012, MNRAS, 420, 340 

\bibitem[Kruijssen et al.(2012)]{2012MNRAS.419..841K} Kruijssen, J.~M.~D., 
Maschberger, T., Moeckel, N., et al.\ 2012, MNRAS, 419, 841 


\bibitem[Kudryavtseva et al.(2012)]{2012ApJ...750L..44K} Kudryavtseva, N., 
Brandner, W., Gennaro, M., et al. 2012, ApJL, 750, L44 


\bibitem[Lada et al.(2000)]{2000AJ....120.3162L} Lada, C.~J., Muench, 
A.~A., Haisch, K.~E., Jr., et al.\ 2000, AJ, 120, 3162 

\bibitem[Lamers et al.(2010)]{2010MNRAS.409..305L} Lamers, H.~J.~G.~L.~M., 
Baumgardt, H., \& Gieles, M.\ 2010, MNRAS, 409, 305 

\bibitem[Lardo et 
al.(2011)]{2011A&A...525A.114L} Lardo, C., Bellazzini, M., Pancino, E., et al.\ 2011, A\&A, 525, A114 













\bibitem[Larsen et al.(2001)]{2001ApJ...556..801L} Larsen, S.~S., Brodie, 
J.~P., Elmegreen, B.~G., et al.\ 2001, ApJ, 556, 801 

\bibitem[Larsen et al.(2008)]{2008MNRAS.383..263L} Larsen, S.~S., Origlia, 
L., Brodie, J., \& Gallagher, J.~S.\ 2008, MNRAS, 383, 263 

\bibitem[Larsen et 
al.(2011)]{2011A&A...532A.147L} Larsen, S.~S., de Mink, S.~E., Eldridge, J.~J., et al.\ 2011, A\&A, 532, A147 


\bibitem[Larsen et 
al.(2012)]{2012A&A...544L..14L} Larsen, S.~S., Strader, J., \& Brodie, J.~P. 2012, A\&A, 544, L14 

\bibitem[Lin 
\& Pringle(1990)]{1990ApJ...358..515L} Lin, D.~N.~C., \& Pringle, J.~E.\ 1990, ApJ, 358, 515 


\bibitem[Longmore et 
al.(2013)]{temp} Longmore, S.N., Kruijssen, J.~M.~D., Bastian, N. et al.~2013, PPVI (in press).


\bibitem[Mackey 
\& Broby Nielsen(2007)]{2007MNRAS.379..151M} Mackey, A.~D., \& Broby Nielsen, P. 2007, MNRAS, 379, 151 

\bibitem[Mackey et al.(2008)]{2008ApJ...681L..17M} Mackey, A.~D., Broby 
Nielsen, P., Ferguson, A.~M.~N., 
\& Richardson, J.~C. 2008, ApJL, 681, L17 

\bibitem[Maraston et 
al.(2004)]{2004A&A...416..467M} Maraston, C., Bastian, N., Saglia, R.~P., et al.\ 2004, A\&A, 416, 467 


\bibitem[Marino et al.(2013)]{2013ApJ...768...27M} Marino, A.~F., Milone, 
A.~P., \& Lind, K.\ 2013, ApJ, 768, 27 


\bibitem[Martell et 
al.(2011)]{2011A&A...534A.136M} Martell, S.~L., Smolinski, J.~P., Beers, T.~C., \& Grebel, E.~K. 2011, A\&A, 534, A136 

\bibitem[McCrady et al.(2005)]{2005ApJ...621..278M} McCrady, N., Graham, 
J.~R., \& Vacca, W.~D.\ 2005, ApJ, 621, 278 


\bibitem[McMillan et al.(2007)]{2007ApJ...655L..45M} McMillan, S.~L.~W., 
Vesperini, E., \& Portegies Zwart, S.~F.\ 2007, ApJL, 655, L45 


\bibitem[Merritt(1985)]{1985AJ.....90.1027M} Merritt, D.\ 1985, AJ, 90, 
1027 


\bibitem[Moeckel 
\& Clarke(2011)]{2011MNRAS.410.2799M} Moeckel, N., \& Clarke, C.~J.\ 2011, MNRAS, 410, 2799 


\bibitem[Monaco et 
al.(2012)]{2012A&A...539A.157M} Monaco, L., Villanova, S., Bonifacio, P., et al.\ 2012, A\&A, 539, A157 




\bibitem[Milone et 
al.(2009)]{2009A&A...497..755M} Milone, A.~P., Bedin, L.~R., Piotto, G., \& Anderson, J. 2009, A\&A, 497, 755 

\bibitem[Moeckel 
\& Throop(2009)]{2009ApJ...707..268M} Moeckel, N., \& Throop, H.~B.\ 2009, ApJ, 707, 268 


\bibitem[Muno et al.(2006)]{2006ApJ...650..203M} Muno, M.~P., Law, C., 
Clark, J.~S., et al. 2006, ApJ, 650, 203 


\bibitem[Mu{\~n}oz et al.(2013)]{2013MNRAS.433.2006M} Mu{\~n}oz, C., 
Geisler, D., \& Villanova, S.\ 2013, MNRAS, 433, 2006 


\bibitem[Olczak et al.(2006)]{2006ApJ...642.1140O} Olczak, C., Pfalzner, 
S., \& Spurzem, R.\ 2006, ApJ, 642, 1140 

\bibitem[Osipkov(1979)]{1979SvAL....5...42O} Osipkov, L.~P.\ 1979, Soviet 
Astronomy Letters, 5, 42 


\bibitem[Pasquini et 
al.(2005)]{2005A&A...441..549P} Pasquini, L., Bonifacio, P., Molaro, P., et al.\ 2005, A\&A, 441, 549 


\bibitem[Peacock et al.(2013)]{2013ApJ...769..126P} Peacock, M.~B., Zepf, 
S.~E., \& Finzell, T. 2013, ApJ, 769, 126 

\bibitem[Piotto et al.(2007)]{2007ApJ...661L..53P} Piotto, G., Bedin, 
L.~R., Anderson, J., et al.\ 2007, ApJL, 661, L53


\bibitem[Portegies Zwart et 
al.(2010)]{2010ARA&A..48..431P} Portegies Zwart, S.~F., McMillan, S.~L.~W., \& Gieles, M. 2010, ARAA, 48, 431 

\bibitem[Prantzos 
\& Charbonnel(2006)]{2006A&A...458..135P} Prantzos, N., \& Charbonnel, C.\ 2006, A\&A, 458, 135 


\bibitem[Renzini(2008)]{2008MNRAS.391..354R} Renzini, A.\ 2008, MNRAS, 
391, 354 

\bibitem[Richer et al.(2013)]{2013ApJ...771L..15R} Richer, H.~B., Heyl, J., 
Anderson, J., et al.\ 2013, ApJL, 771, L15 

\bibitem[Rogers 
\& Pittard(2013)]{2013MNRAS.431.1337R} Rogers, H., \& Pittard, J.~M.\ 2013, MNRAS, 431, 1337 


\bibitem[Rossa et al.(2006)]{2006AJ....132.1074R} Rossa, J., van der Marel, 
R.~P., B{\"o}ker, T., et al. 2006, AJ, 132, 1074 


\bibitem[Salpeter(1955)]{1955ApJ...121..161S} Salpeter, E.~E.\ 1955, ApJ, 
121, 161 

\bibitem[Sana et 
al.(2010)]{2010A&A...515A..26S} Sana, H., Momany, Y., Gieles, M., et al.\ 2010, A\&A, 515, A26 

\bibitem[Sana et al.(2012)]{2012Sci...337..444S} Sana, H., de Mink, S.~E., 
de Koter, A., et al.\ 2012, Science, 337, 444 

\bibitem[Sbordone et 
al.(2011)]{2011A&A...534A...9S} Sbordone, L., Salaris, M., Weiss, A., \& Cassisi, S.\ 2011, A\&A, 534, A9 


\bibitem[Scally 
\& Clarke(2001)]{2001MNRAS.325..449S} Scally, A., \& Clarke, C.\ 2001, MNRAS, 325, 449 


\bibitem[Seale et al.(2012)]{2012ApJ...751...42S} Seale, J.~P., Looney, 
L.~W., Wong, T., et al.\ 2012, ApJ, 751, 42 

\bibitem[Seth et al.(2008)]{2008ApJ...687..997S} Seth, A.~C., Blum, R.~D., 
Bastian, N., Caldwell, N., \& Debattista, V.~P. 2008, ApJ, 687, 997 

\bibitem[Schweizer 
\& Seitzer(1998)]{1998AJ....116.2206S} Schweizer, F., \& Seitzer, P.\ 1998, AJ, 116, 2206 


\bibitem[Siess et 
al.(2000)]{2000A&A...358..593S} Siess, L., Dufour, E., \& Forestini, M.\ 2000, A\&A, 358, 593 


\bibitem[Sills 
\& Glebbeek(2010)]{2010MNRAS.407..277S} Sills, A., \& Glebbeek, E.\ 2010, MNRAS, 407, 277 

\bibitem[Smith et al.(2007)]{2007AJ....134..846S} Smith, N., Bally, J., 
\& Walawender, J.\ 2007, AJ, 134, 846 

\bibitem[Smith et al.(2013)]{2013MNRAS.429.1324S} Smith, N., Arnett, W.~D., 
Bally, J., Ginsburg, A., \& Filippenko, A.~V.\ 2013, MNRAS, 429, 1324 

\bibitem[Spezzi et al.(2012)]{2012MNRAS.421...78S} Spezzi, L., de Marchi, 
G., Panagia, N., Sicilia-Aguilar, A., 
\& Ercolano, B.\ 2012, MNRAS, 421, 78 


\bibitem[Throop 
\& Bally(2008)]{2008AJ....135.2380T} Throop, H.~B., \& Bally, J.\ 2008, AJ, 135, 2380 


\bibitem[Vesperini et al.(2010)]{2010ApJ...718L.112V} Vesperini, E., 
McMillan, S.~L.~W., D'Antona, F., \& D'Ercole, A.\ 2010, ApJL, 718, L112 

\bibitem[Villanova 
\& Geisler(2011)]{2011A&A...535A..31V} Villanova, S., \& Geisler, D.\ 2011, A\&A, 535, A31 

\bibitem[Weidenschilling(1977)]{1977Ap&SS..51..153W} Weidenschilling, S.~J.\ 1977, Ap\&SS, 51, 153 


\bibitem[Yang et al.(2013)]{2013arXiv1304.5865Y} Yang, W., Bi, S., Meng, 
X., \& Liu, Z.\ 2013, ApJ, submitted (arXiv:1304.5865)


\bibitem[Yusof et al.(2013)]{2013MNRAS.433.1114Y} Yusof, N., Hirschi, R., 
Meynet, G., et al.\ 2013, MNRAS, 433, 1114 



\end{thebibliography}
\end{document}